\documentclass[useAMS,usenatbib,usegraphicx]{mn2e}
\usepackage{times}

\title[Trends of galaxy age in A901/902]{Morphology-dependent trends 
 of galaxy age with environment in Abell 901/902 seen with COMBO-17}

\author[C. Wolf et al.]{C. Wolf$^{1}$, M. E. Gray$^{2}$,
	A. Arag\'on-Salamanca$^{2}$, K. P. Lane$^{2}$,
	K. Meisenheimer$^{3}$ \smallskip \smallskip \\
$^{1}$Dept.\ of Physics, University of Oxford, Keble Road, Oxford, OX1 3RH,
U.K., email: cwolf@astro.ox.ac.uk\\
$^{2}$School of Physics and Astronomy, The University of Nottingham, University Park, Nottingham, NG7 2RD, U.K.\\
$^{3}$Max-Planck-Institut f\"ur Astronomie, K\"onigstuhl 17, D-69117 
Heidelberg, Germany}

\begin{document}
\date{\today}
\maketitle

\begin{abstract}
We investigate correlations between galaxy age and environment in the Abell 
901/2 supercluster for separate morphologies. Using COMBO-17 data, we define 
a sample of 530 galaxies, complete at $M_V -5\log h<-18$ on an area of 
$3.5\times 3.5$~(Mpc/$h$)$^2$. We explore several age indicators including 
an extinction-corrected residual from the colour-magnitude relation (CMR). 
As a result, we find a clear trend of age with density for galaxies of all 
morphologies that include a spheroidal component, in the sense that galaxies 
in denser environments are older. This trend is not seen among Scd/Irr 
galaxies since they all have young ages. However, the trend among the 
other types is stronger for fainter galaxies. While we also see an expected
age-morphology relation, we find no evidence for a morphology-density 
relation at fixed age.
\end{abstract}


\section{Introduction}
One of the long-standing mysteries of galaxy evolution is the origin of the morphology-density relation \citep{D80}: high-density environments such as 
galaxy clusters are dominated by spheroidal galaxies while disk galaxies are 
the main population in the lower density field. This trend is also reflected 
in a colour-density relation \citep[e.g.][]{B99}: High-density environments 
are dominated by old stellar populations, not only because the first galaxies 
started to collapse there, but also because star formation seems effectively 
suppressed still today \citep[e.g.][]{Ko01,Lew02,Go03,G04,Po06}.

\citet{K04} found strong correlations between stellar age, morphology and 
the density of the local galaxy environment using the spectroscopic sample 
from SDSS DR2. They show quantitatively how stellar age increases with more 
spheroidal morphologies and towards denser environments. More recently, 
\citet{Ba06} analyzed a sample of $\sim$80,000 galaxies from SDSS DR4, and
report that no residual morphology-density relation is seen at fixed colour.
Conversely, a strong colour-density relation still remains at fixed morphology.

Colour alone provides only a crude estimate of stellar age as demonstrated by 
the colour-magnitude relation (CMR) of elliptical, passively evolving galaxies 
\citep{Ko98}: the slope of the CMR reflects a metallicity sequence with 
luminosity or stellar mass, while the offset of an individual galaxy from the 
CMR (its CMR residual) could reflect its stellar age. Of course, dust 
reddening affects the colour further, especially in the cluster Abell 901/902, 
which has a high fraction of dust-reddened galaxies on the red sequence 
\citep{WGM05}.

In this paper we investigate these correlations in a single dense 
environment, that of the clusters Abell 901/902 observed by COMBO-17. The 
clusters are arranged in a very complex structure with filaments and four major 
concentrations. The hot gas content as seen in X-rays as well as the galaxy 
velocities suggest a highly dynamic environment that is far from virialized 
(Gray et al., in preparation). 
It may be a consequence of this action, that the cluster shows 
an unexpected, rich population of dusty star-forming galaxies accounting 
for more than a third of the red sequence. 

In Sect.~2 we present the galaxy sample, and in Sect.~3 we discuss various 
SED-based stellar age indicators. Correlations between stellar age, galaxy 
morphology and local galaxy density are analysed in Sect.~4. In 
Sect.~5 we mention possible caveats of our study and finally discuss our 
results in Sect.~6. Throughout the paper, we use Vega magnitudes and the 
cosmological parameters $(\Omega_{\rm m},\Omega_\Lambda) =(0.3,0.7)$ and 
$H_0=h\times 100$~km/(s~Mpc).

\begin{figure*}
\centering
\includegraphics[clip,angle=270,width=0.67\hsize]{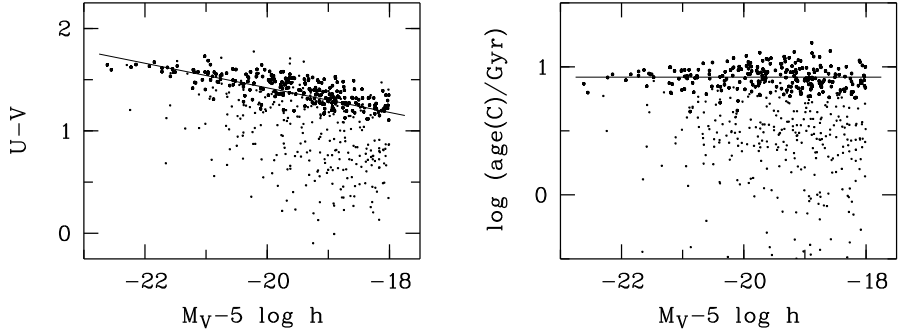}
\caption{{\it Left:} Colour-magnitude diagram of the A901/902 supercluster
with CMR fit. Typical errors in the colour are 0.03~mag (statistical) with
an additional 0.03~mag uncertainty in the zeropoint from relative passband 
calibration. Dust-poor ($E_{B-V}<0.1$) red-sequence galaxies are shown as 
large points. {\it Right:} Inferred age-density diagram using indicator (C).
\label{CMD}}
\end{figure*}

\section{Data}

This work is based on the data of the Abell 901/902 supercluster field in the 
COMBO-17 survey \citep[see][for a general survey characterization]{W04}. We 
start from the conservative cluster member sample defined by \citet{WGM05}, 
which contains 795 galaxies in a $\sim 30\arcmin \times 30\arcmin$ field. The 
sample was selected to have photometric redshifts in the range $z=[0.155,0.185
]$ and luminosities brighter than $M_V=-17$. At $M_V<-18.5$, this sample is 
$>98$\% complete, owing to a photometric redshifts accuracy of $\sigma_z<0.01$.
The accuracy was measured with 249 spectroscopic redshifts of cluster members 
obtained with 2dF. This master sample was used to define environment in the 
form of projected surface densities $\Sigma_{10}$ in units of $({\rm Mpc}/h)
^{-2}$. They stem from a wide velocity interval of $\pm 4500$~km/s required by
the photo-z errors, although the cluster occupies a truly much smaller volume 
along the line-of-sight. Hence, the $\Sigma_{10}$ values resemble much larger 
space densities, than a comparison with field densities in similarly large 
velocity intervals would suggest.

For these galaxies, we then determined visual morphological types on the deep
R-band image of COMBO-17, that is co-added from 20,300~s of exposure with WFI
at the ESO/MPI 2.2-m-telescope at La Silla, Chile. The image is characterized 
by a $0\farcs8$ PSF and a point-source detection limit of $R_{5\sigma}\approx 
26$. At the spectroscopic distance of this cluster $z=0.165$, an angle of 
$1\arcsec$ corresponds to a projected physical separation of $2.0/h$~kpc.
We found that a reliable morphological classification was only possible for 
objects of $M_V<-18$, yielding a restricted sample of 530 galaxies. 
An average galaxy at $M_V=-18$ has a visible diameter of $\ga 2\arcsec$ 
($\la 3\times$PSF), which may be the real reason for the limit. 

The
morphological classification was carried out by three independent authors (AAS,
MEG, KL) and is described in more detail by \citet{Lane06}. For the purposes
of this paper, four morphological classes are considered, which are E, S0, 
Sa/b, and Scd/Irr. While the spiral classes are broadly defined due to small
object numbers, the spheroid classes try hard to discriminate E's from S0's 
although differences in their appearance are small and orientation-dependent.

This final sample is $>95$\% complete in terms of cluster members. A similarly
defined field sample from COMBO-17 suggests, that statistically 37 out of the 
530 galaxies are field contaminants (mostly blue spirals), arising from the 
wide velocity interval used for the sample definition. However, no correction 
for field contamination is attempted in this work, as it is particularly low 
among spheroids, which are at the focus of our attention.

\begin{figure}
\centering
\includegraphics[clip,angle=270,width=0.7\hsize]{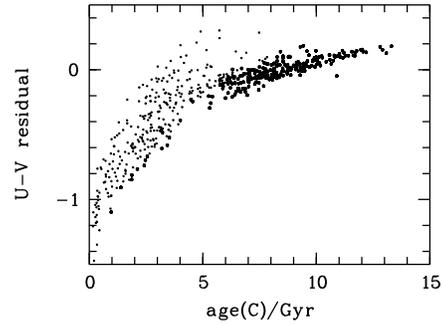}
\caption{CMR-corrected age compared to CMR residual for dust-poor galaxies
(large points, $E_{B-V}<0.1$) and all other galaxies (small points).
\label{ageBC}}
\end{figure}

\begin{figure*}
\centering
\includegraphics[clip,angle=270,width=0.85\hsize]{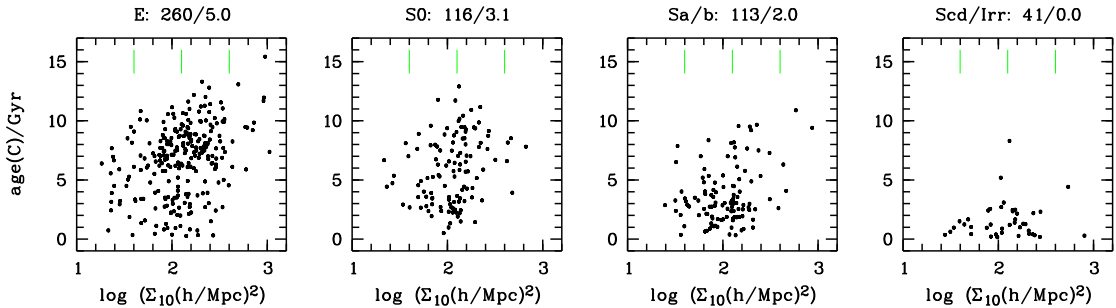}
\caption{Age-density relation at fixed morphology in A901/902. 
Number of objects and Student's $t$ statistic from a Spearman rank test are 
listed above the panels. The null hypothesis of no correlation between age 
and density is rejected at $>99.9$\%, $>99$\% and $\sim 95$\% for E, S0 and 
Sa/b galaxies, respectively, but accepted for Scd/Irr galaxies. Grey lines
separate the density bins for the left panel of Fig.~5.
\label{ADrel}}
\end{figure*}

\begin{figure*}
\centering
\includegraphics[clip,angle=270,width=0.85\hsize]{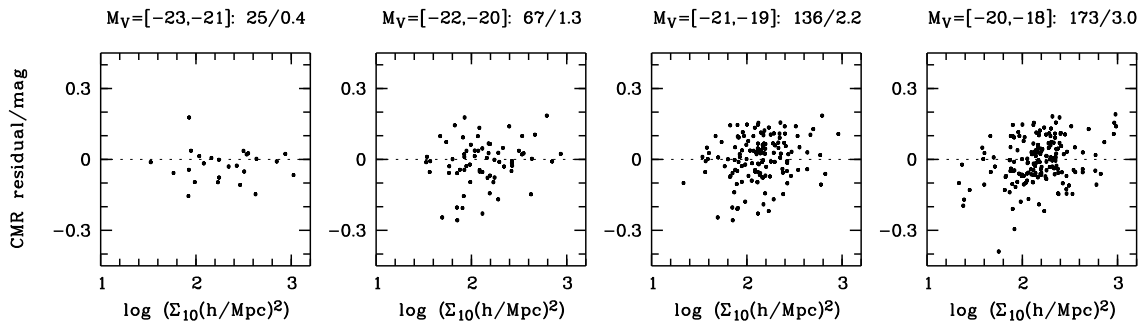}
\caption{CMR residuals vs. density: Dust-free galaxies of all types are split 
into (overlapping, for better statistics) luminosity bins. No correlation is 
accepted at high luminosity but increasingly rejected towards lower $M_V$ at 
levels of $<50$\%, $\sim 80$\%, $>95$\% and $>99$\%. Number of objects and 
Student's $t$ statistic from a Spearman rank test are given. The errors of 
individual CMR residuals are 0.03~mag, while CMR slope errors mostly shift 
the zeropoints of each bin.
\label{ADrel2}}
\end{figure*}

\section{Colour-magnitude relation and stellar age indicators}

The COMBO-17 catalogue provides restframe $UBV$ magnitudes obtained by 
integrating the SED of the best-fitting template over
the desired bandpasses. A restframe colour-magnitude diagram (CMD)
is thus obtained for our sample, where the red sequence and blue cloud can be
readily identified. Statistical errors in the restframe colour index $U-V$ are
$\sim 0.03$~mag, and the zeropoint is uncertain by another $\sim 0.03$~mag 
resulting from uncertainties in the relative bandpass calibration. COMBO-17 
colours represent a fixed-size aperture outside the atmosphere 
with identical spatial weighting in all bands. They are thus not representative 
of total object colours. The COMBO-17 classification estimates photometric
redshifts, but also a measure of age of the stellar population and the amount of 
dust reddening \citep[for details see][]{WGM05}. In Fig.~\ref{CMD} we show the 
CMD with dust-poor (estimated $E_{B-V}<0.1$) red-sequence galaxies emphasized
as large points. A linear fit to the red sequence yields a colour-magnitude 
relation (CMR) of

\begin{equation}
	(U-V)_{\rm rest} = 1.45 - 0.10 (M_V-5\log h +20)	~,
\end{equation}
which fits in well with a redshift extrapolation of the evolving red sequence
of field galaxies in COMBO-17 \citep{B04}. Here, the slope is slightly steeper
than usual (0.10 instead of 0.08), which may be a luminosity-dependent colour 
bias introduced by the fixed-size aperture photometry in COMBO-17. However, 
in the absence of internal colour gradients in galaxies this slope would be 
identical for both aperture and total galaxy colours.

We wish to study relations between stellar age and other galaxy properties, 
and use more advanced indicators than restframe colours, which are clearly 
affected by metallicity and dust reddening. We aim for an age ranking rather
than absolute age determination. Also, the luminosity-weighted SEDs reflect 
more the time passed since the last major star formation event instead of a 
mass-weighted mean age. The formal COMBO-17 template fits to the observed SEDs
will assist us in defining the colour/age measures. These templates are a time 
sequence after a burst of star formation assuming a fixed starting metallicity 
and an exponentially declining star-formation rate (SFR) with $\tau=1$~Gyr. 
They include dust reddening from an SMC law ranging over $E_{B-V}=[0.0,0.5]$. 
This age is a rough estimate on a possibly wrong scale, but age ranking would 
be recognized correctly, if galaxies of equal metallicity and star formation
history (SFH) were considered. The present work is 
mostly concerned with spheroidal galaxies, where the bulk of the stellar 
population is old and large variations in SFH are not expected. We consider 
three different measures of galaxy colour or age of the stellar population:

{\bf A: The formal template age from the COMBO-17 fit.} This age corresponds 
directly to the restframe colour which the galaxy would expose if its dust was
removed. However, it ignores variations in metallicity and the details of 
the SFH of an individual galaxy. With the CMR being a metallicity sequence, 
this measure is not satisfactory for our purposes. 

{\bf B: The CMR residual.} This is a pure colour offset of galaxy $i$ from 
the CMR at its luminosity $(U-V)_i-(U-V)_{\rm CMR}(M_V,i)$. Various works 
have suggested that the deviation of galaxy colour from the CMR at fixed 
luminosity, mass or velocity dispersion, is an age indicator \citep[for a 
discussion, see][]{B05}. This should work well for dust-free galaxies, where 
age drives colour when metallicity is fixed. However, the surprisingly large 
fraction of dusty red-sequence galaxies in A901/902 ($\sim 35$\%) makes this 
approach only viable for part of our early-type sample.

{\bf C: A dust-corrected CMR residual} or {\bf a CMR-corrected template age}. 
This new measure is a combination of the two previous ones and alleviates the 
insensitivity of (A) to metallicity and the insensitivity of (B) to dust.
It can either be expressed as the CMR residual a galaxy would have after 
taking away its dust. Or it can be expressed as the template age obtained 
by subtracting the colour offset $\delta (U-V)_{CMR}=-0.10(M_V-5\log h+20)$ 
introduced by the CMR slope for each galaxy, keeping the known dust content
fixed, and obtaining the age that corresponds to the corrected colour. The
advantage of this approach is that it allows to take into account all three 
factors driving galaxy colour, age, metallicity and dust, albeit only at a 
crude level.

In the following, we will discuss relations between age, density and 
morphology based on age indicator (C). Indicator (B) produces very similar 
results as long as is not used with dusty galaxies, where it is a-priori not
expected to work. This point is emphasized by Fig.~\ref{ageBC} that plots 
CMR residual (B) against CMR-corrected age (C). The correlation between the 
two indicators is tight among dust-poor galaxies, but widens considerably for
dustier galaxies.

\begin{figure*}
\centering
\includegraphics[clip,angle=270,width=0.3\hsize]{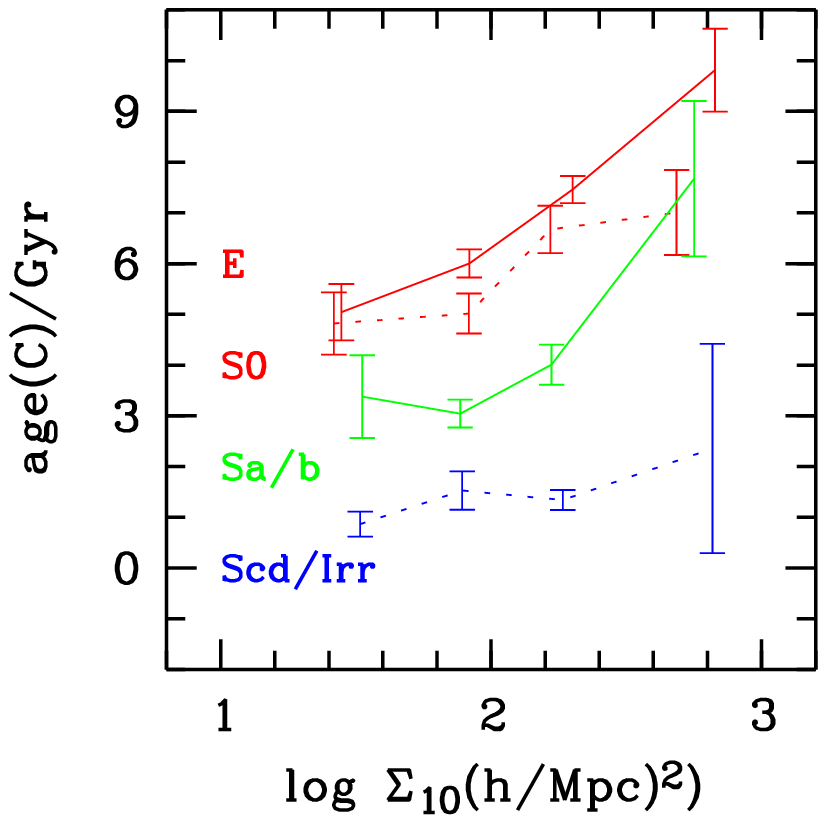} \includegraphics[clip,angle=270,width=0.3\hsize]{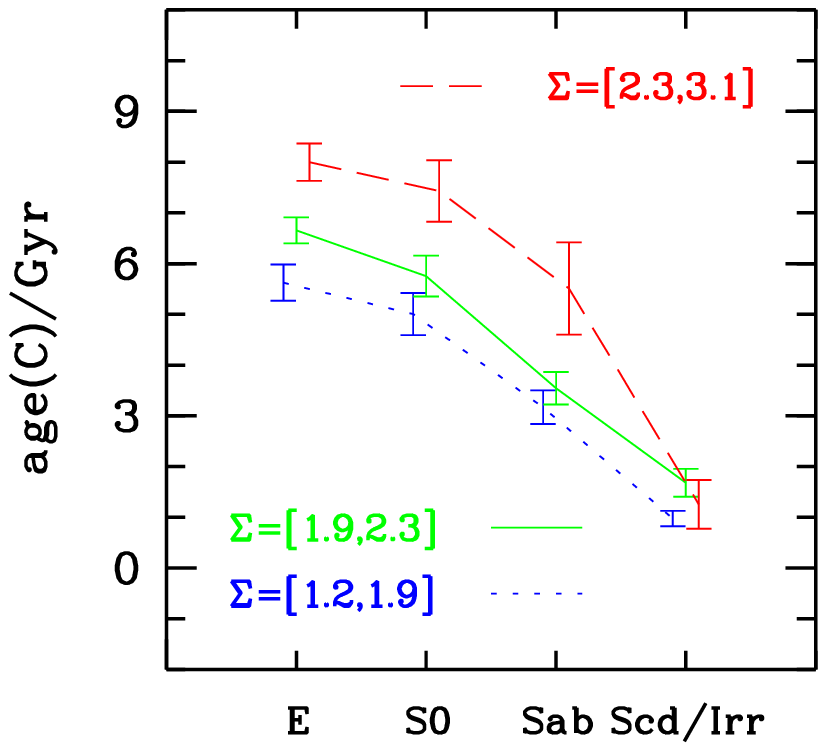} \includegraphics[clip,angle=270,width=0.3\hsize]{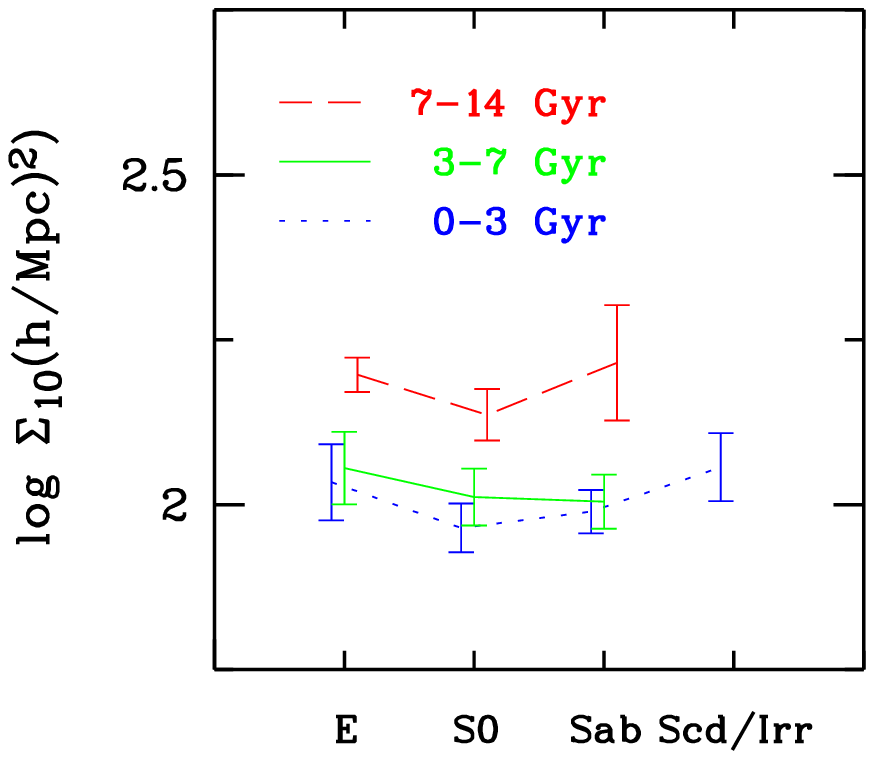}
\caption{2-parameter relations at fixed third parameter in the A901/902: 
There is a trend of age with density at fixed morphology ({\it left panel}) 
and a prominent age-morphology relation even at fixed density ({\it center 
panel}), but there is no evidence for a morphology-density relation at fixed 
age ({\it right panel}).
\label{2Drels}}
\end{figure*}

\section{Age-morphology-density relations}

Age and density are continuous variables, while morphology is described in 
class bins, which can be ranked according to the prominence of the
spheroidal component. In principle, morphology could be expressed as a 
continuous variable, if we used quantitative morphological descriptions.

We first investigate a plot of age vs. density at fixed morphologies (see 
Fig.~\ref{ADrel}), which in principle contains already the full 
information. For the galaxy types from E to Sb we find a correlation in the 
sense that galaxies in dense environments are on average older. They are 
confirmed by Spearman rank tests at $>99$\% significance for both E's and 
S0's, although the situation is less convincing to the eye for S0's. For Sa/b 
galaxies we find that a null correlation is rejected at $>95$\% significance. 
Scd/Irr galaxies show no signs of any age-density relation.

 Some authors have found age variations with mass along the red sequence
 \citep[and references therein]{C03,C06}, which would make our age
 ranking only hold at fixed galaxy mass. However, high-mass galaxies are
 then preferentially older and also found at higher-density, such that 
 the true trend might be stronger than our result suggests.

While the Spearman rank correlation test makes no assumptions on the form of
the data distributions, we still wish to point out that most galaxies in our
sample are clustered around a density of 120 galaxies per (Mpc/$h$)$^2$ area.
The densest regions with at $\Sigma_{10}>300$/(Mpc/$h$)$^2$ cover only a small
area,
 as they correspond typically to a clustercentric radius of $1\farcm0$
 or $120/h$~kpc. Also, our field is not large enough to reach out to 
 densities $\Sigma_{10}<20$/(Mpc/$h$)$^2$, compared to the COMBO-17 random 
 field with $\Sigma_{10}\approx 4.7$.
However, these selection effects on density are entirely
independent of galaxy age and morphology, because our sample is purely limited
by luminosity and volume and complete with respect to other galaxy properties.

In Fig.~\ref{ADrel2} we split the population by luminosity 
instead of morphology. Also, we plot the measured CMR residual rather than
the derived age, but then only for dust-poor galaxies. At high luminosity 
(brighter than $M_V^*$, i.e. $-20.5$ for red galaxies in COMBO-17)
we find no indication of a trend with environment. However, towards lower $M_V$
we find an increasing significance for a trend in the sense, that galaxies in
high-density environments have more positive CMR residuals than those in lower 
densities. Thus, at high-density faint galaxies have redder CMR residuals than 
bright galaxies, while at low densities they have bluer residuals, keeping the 
mean CMR in the assumed place. An interesting corollary of this observation is
a subtle environmental dependence in the slope of the CMR, to be explored in a 
forthcoming paper.

We now look at all possible 2-parameter correlations while keeping the third 
variable fixed. Fig.~\ref{2Drels} shows them as mean quantities in one variable 
after binning the other two. First we look again at the age-density relation 
but this time in bins of morphological type and density. As before, we find a 
clear increase in age with density for E and S0 galaxies. There is a less 
significant trend for Sa/b spirals and no trend for Scd/Irr galaxies. The center 
panel of Fig.~\ref{2Drels} shows a clear age-morphology relation at fixed 
density for all density bins. The age increases towards earlier type galaxies, 
from 1--2~Gyr for Scd/Irr's to 5--8~Gyr for E and S0 galaxies. In fact, at 
all densities we find S0's to be slightly younger than E's. Finally, in the 
right panel we inspect a possible morphology-density relation at fixed age, 
but find no evidence for it. In all age bins, the mean density of the sample 
is practically independent of morphology.

\section{Caveats}

Our sample contains 530 galaxies, which are selected by photometric redshift 
and luminosity, and is more than 95\% complete. The completeness is independent 
of stellar age, morphology or environment (surface density), three measures of 
galaxy properties with possible correlations among them. Observed correlations 
should thus not be the result of sample selection effects. 

Measurements of all three galaxy properties are subject to statistical and 
possibly systematic errors. However, we believe that these can not lead to 
the presence or absence of the observed relations. The relations are clearly
observed with both stellar age indicators (B) and (C). The indicators do not 
precisely measure mean stellar age but rather time passed since the last major 
star-formation event. Even then the absolute age scale is uncertain, while the 
age ranking should be reliable. A study by \citet{B05} appears to rule out that 
CMR residuals are dominated by metallicity effects and confirms their 
interpretation as age effects. 
 \citet{T00} suggest that CMR residuals in age and metallicity are
 anti-correlated at fixed mass, implying a larger underlying age scatter
 than assumed from the colour residuals. In this case, we would observe a
 reduced signal from a truly stronger effect.

The morphology assessed by human classifiers from groundbased images is not 
reproducible. However, averaging the classifications from three persons allows 
to quantify and reduce noise. The weakest point is probably the discrimination 
between E and S0 galaxies. However, even mixing the bins for these two types
would not remove any of the observed trends and relations.

On the whole, we believe that any possible systematic biases in morphology or 
age should be independent of each other and independent of environment. 
However, we have to guard ourselves against biases from luminosity-dependent 
galaxy properties such as metallicity, or in the presence of internal colour 
gradients also the fraction of light inside our fixed-size aperture defining 
the SED. When looking at the relation between density and the CMR residual 
in luminosity bins we eliminated these problems.

\section{Discussion}

We have observed a clear trend of increasing age with increasing density of the 
environment for galaxies with spheroidal components in the cluster A901/902. 
The significance of this age-density trend increases when going to fainter 
galaxies, while it is invisible at $M_V<-21$. In contrast, we found no 
evidence for any morphology-density relation at fixed galaxy age.

\citet{K04} investigated relations between environment, morphology, SFH and 
mass of galaxies in the SDSS. Their sample represents a random galaxy field 
with their highest density bin corresponding to $\Sigma_{10}\ge 2$/(Mpc/$h
$)$^2$ on our scale, which is a factor of ten below our lowest bin. From this
they found very much the same relations as we do in the cluster: 
a declining star-formation rate and an increasing stellar 
age with increasing density are the strongest relation in place. A trend 
for older galaxies to be more spheroid-dominated is evident from their and 
our data, but has become textbook knowledge a while ago. Also, they find no 
evidence for an explicit morphology-density relation at fixed mass.
 
\citet{Ha06} also find a clear age-density relation in their analysis of the 
supercluster Abell 2199 from the SDSS DR4 data set. However, their result 
applies to the full galaxy population and is not differentiated by type. They observe, that fainter galaxies are generally younger and that the critical 
density for the marked increase in age is higher for fainter galaxies. 
It is difficult to compare our results directly to theirs, because we have 
no reliable absolute age measures and have deliberately eliminated the CMR 
slope from our age estimator. However, we still agree on the finding that 
faint galaxies have younger ages at lower density than at high density.

Also, \citet{Sm06} have obtained strong evidence for an age trend with 
clustercentric radius among red-sequence cluster galaxies and suggest that 
the trend may be stronger for lower-mass galaxies. In our complex cluster
environment, \citet{Lane06} have compared trends of galaxy properties with 
environment using several measures of the latter: they suggest that the
underlying relations are with local galaxy density, such as $\Sigma_{10}$ 
or dark-matter density, rather than clustercentric radius. Of course, in 
virialized clusters of regular shape one could not observe any difference 
between the two environmental measures. From our data set, we can confirm the 
age trend with local density. We have further identified that in Abell 901/2 
this trend applies independently to E and S0 galaxies, and probably to spiral 
galaxies as well. Thus, the trend is not a simple result of the morphological 
mix changing with environment, but affects galaxies of all morphologies. We 
also confirm the strength of this trend to increase with decreasing luminosity.

\citet{Ne05} have also reported an increased age scatter among lower-mass 
red-sequence galaxies in clusters. The right panel of our Fig.~1 shows the 
same result for our dust-poor red-sequence galaxies, an increased age scatter 
with lower luminosity. We note that this scatter is not a result of noise in 
our sample (that should be at the level of 0.03~mag). This is also consistent 
with the stronger age-density trend among fainter galaxies as seen in the 
right panel of our Fig.~\ref{ADrel2}. 

Statistically, the current density in the environment of a galaxy is 
correlated with the density at its birth. Thus, galaxies which are today 
in high density environments started their lives mostly in high density 
environments, regardless of their morphology. Hierarchical models of galaxy 
formation \citep[e.g.][]{DL06} also indicate that galaxies in high-density 
regions started their lives (and forming stars) earlier and were exposed for a 
longer time to the environmental effects that could quench star formation, 
implying older average ages. Altogether, these results suggest that the 
primary relation among the parameters is the age-density relation. Age is 
also correlated with morphology, as old galaxies had more time to undergo 
interactions or mergers leading to bulge formation and/or the cessation of 
star formation.

Massive red-sequence galaxies show no significant age trend in our relatively 
high-density environment, while evidence for field ellipticals being younger 
than cluster ellipticals \citep[e.g.][]{Ku02} suggests an age trend appearing 
at densities below the regime of our study. In contrast, low-mass galaxies 
show an age trend across our higher-density regime, suggesting a mass 
dependence of the density regime 
in which galaxies lose their ability to form stars. This mass dependence is 
consistent with a picture in which massive galaxies terminate their star 
formation via feedback, while less massive galaxies have their SF suppressed 
by a dense environment. Future surveys of super-cluster environments would 
help to firm up these trends and determine their physical origin.

\vspace*{-0.5cm}

\section*{acknowledgements}
CW was supported by a PPARC Advanced Fellowship, KPL by a PPARCstudentship 
and MEG by an Anne McLaren Research Fellowship from the University of 
Nottingham. We thank H. Kuntschner and S. D. M. White for discussions
and the referee, Russell Smith, for valuable comments improving the manuscript.

\end{document}